\documentclass[runningheads]{llncs}
\usepackage{graphicx}
\usepackage{hyperref}
\usepackage{listings}
\usepackage{todonotes}
\usepackage{tikz}
\usepackage{pgfplots}
\usepackage{multirow}

\pgfplotsset{compat=1.15}
\usetikzlibrary{patterns}

\lstdefinestyle{c_code_style}{
basicstyle=\small\ttfamily,
keywordstyle=\color{blue},
commentstyle=\color{gray},
breaklines=true,
language=C,
morekeywords={uint64_t},
}

\newcommand{\OpenMP}{\textsc{OpenMP}}
\newcommand{\Summit}{\textsc{Summit}}

\AtBeginDocument{

}

\begin{document}
\title{Experience Report: Writing A Portable GPU Runtime with \OpenMP{} 5.1}
%
%
\author{Shilei Tian\inst{1}\orcidID{0000-0001-6468-6839} \and
Jon Chesterfield\inst{2}\orcidID{0000-0002-8546-2014}\and
Johannes Doerfert\inst{3}\orcidID{0000-0001-7870-8963} \and
Barbara Chapman\inst{1}\orcidID{0000-0001-8449-8579}}


\institute{
Department of Computer Science, Stony Brook University, USA \\
\email{\{shilei.tian, barbara.chapman\}@stonybrook.edu}
\and
Advanced Micro Devices, UK \\
\email{jchester@amd.com}
\and
Mathematics and Computer Science, Argonne National Laboratory, USA \\
\email{jdoerfert@anl.gov}
}
\maketitle              

\setcounter{footnote}{0}

\begin{abstract}
GPU runtimes are historically implemented in CUDA or other vendor specific languages dedicated to GPU programming.
In this work we show that \OpenMP{} 5.1, with minor compiler extensions, is capable of replacing existing solutions without a performance penalty.
The result is a performant and portable GPU runtime that can be compiled with LLVM/Clang to Nvidia and AMD GPUs without the need for CUDA or HIP during its development and compilation.

While we tried to be \OpenMP{} compliant, we identified the need for compiler extensions to achieve the CUDA performance with our \OpenMP{} runtime.
We hope that future versions of \OpenMP{} adopt our extensions to make device programming in \OpenMP{} also portable across compilers, not only across execution platforms.

The library we ported to \OpenMP{} is the \OpenMP{} device runtime that provides \OpenMP{} functionality on the GPU.
This work opens the door for shipping \OpenMP{} offloading with a Linux distribution's LLVM package as the package manager would not need a vendor SDK to build the compiler and runtimes.
Furthermore, our \OpenMP{} device runtime can support a new GPU target through the use of a few compiler intrinsics rather than requiring a reimplementation of the entire runtime.

\end{abstract}

\section{Introduction}
In this paper, we describe how we ported the LLVM \OpenMP{} device runtime library to \OpenMP{} 5.1 using only minor extensions not available in the standard.
The \OpenMP{} device runtime provides the \OpenMP{} functionalities to the user and implementation code on the device, which in this context means on the GPU.
As an example, it provides the \OpenMP{} API routines as well as routines utilized by the compiler e.g., for worksharing loops.

Our work replaced the original LLVM \OpenMP{} device runtime implemented in CUDA
to allow for code reusibility between different targets, e.g. AMD and Nvidia.
It further lowers the bar to entry for future targets that only need to provide a few target specific intrinsics and minimal glue code.

The \OpenMP{} device runtime library can now be shipped with pre-build LLVM packages as they only need LLVM/Clang to build it; neither a target device nor vendor SDKs are required, which lowers the barrier to entry for \OpenMP{} offloading.
This work is a proof of concept for writing device runtime libraries in \OpenMP{}, with identical functionality and performance to that available from CUDA or HIP compiled with the same LLVM version.

The remainder of the paper is organized as follows.
We discuss background and motivation in Section~\ref{sec:background}.
Section~\ref{sec:implementation} presents our approach, which is followed by an evaluation in Section~\ref{sec:evaluation}.
Finally, we conclude the paper in Section~\ref{sec:conclusion}.

\section{Background}
\label{sec:background}
When compiling from one language to another, there are usually constructs that are straightforward in the former and complicated or verbose in the latter.
For example, a single \OpenMP{} construct \lstinline{#pragma omp parallel for} is lowered into a non-trivial amount of newly introduced code in the application, including calls into a runtime that provides certain functionality, like dividing loop iterations.
In this work, the input is \OpenMP{} target offloading code, that is the \OpenMP{} target directive and the associated code, and the output is ultimately Nvidia's PTX or AMD's GCN assembler.

\subsection{Device Runtime Library}
The LLVM \OpenMP{} device runtime library contains the various functions the compiler needs to implement \OpenMP{} semantics when the target is an Nvidia or AMD GPU.
It is basically \lstinline{libomp} for the GPU.
The original implementation in LLVM was in CUDA~\cite{Jacob17}, compiled with Nvidia's NVCC to PTX assembler which was linked with the application code to yield a complete program.
We later adapted that source to compile alternatively as HIP, which is close enough to CUDA syntax for the differences to be worked around with macros.
Prior to this work the device runtime was hence comprised of sources in a common and target dependent part.
In order to let the target dependent compiler recognize the code, target dependent keywords (such as \lstinline{__device__} and \lstinline{__shared__} in CUDA) are replaced with macros (\lstinline{DEVICE} and \lstinline{SHARED}), and the header where these macros are defined will be included accordingly depending on the target.
The basic idea is visualized in in \autoref{lst:macros_in_devicertls}.

\begin{lstlisting}[style=c_code_style, caption={Macros in current device runtime.}, label={lst:macros_in_devicertls}]
// Common part
DEVICE void *__kmpc_alloc_shared(uint64_t bytes);
SHARED int shared_var;
// CUDA header
#define DEVICE __device__
#define SHARED __shared__
// AMDGCN header
#define DEVICE __attribute__((device))
#define SHARED __attribute__((shared))
\end{lstlisting}

This strategy works.
For Nvidia offloading the source is compiled as CUDA, for AMDGPU offloading it is compiled as HIP.
Both produce LLVM bitcode but with different final targets, Nvidia's PTX and AMD's GCN respectively.
However, if a programming model does not adequately resemble CUDA, such as OpenCL or Intel's DPC++~\cite{intel-dpcpp}, the approach will become less straight forward.

What's more, this setup requires vendor SDKs (such as CUDA Toolkit or ROCm Developer Tools) to compile the device runtime, which creates a barrier for the package managers of Linux distributions.
In practice that means the LLVM \OpenMP{} installed from Linux distributions does not support offloading out of the box because the package would require a dependence on the CUDA or ROCm package, among other things.

\subsection{Compilation Flow of \OpenMP{} Target Offloading in LLVM/Clang}\label{sec:compilation-flow}
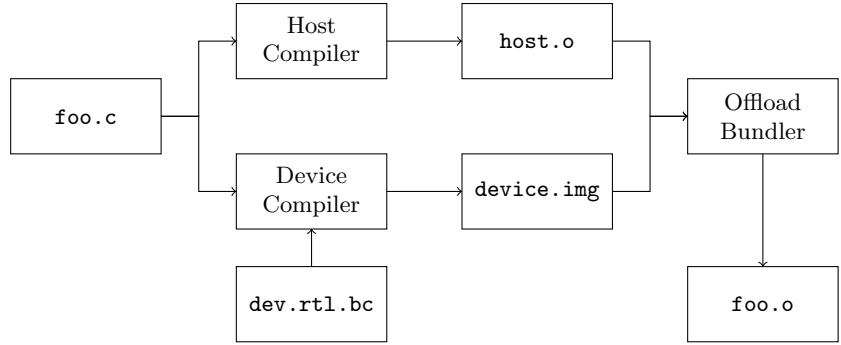
\begin{figure}[hbt!]
\centering
\begin{tikzpicture}[scale=1]
\draw (0,0) rectangle (2,1);
\node[align=center] at (1,0.5) {\lstinline{foo.c}};

\draw[->] (2,0.5) -- (2.5,0.5) -- (2.5,1.5) -- (3,1.5);

\draw (3,1) rectangle (5,2);
\node[align=center] at (4,1.5) {Host\\Compiler};

\draw[->] (2,0.5) -- (2.5,0.5) -- (2.5,-0.5) -- (3,-0.5);

\draw (3,-1) rectangle (5,0);
\node[align=center] at (4,-0.5) {Device\\Compiler};

\draw[->] (5,1.5) -- (6,1.5);

\draw (6,1) rectangle (8,2);
\node[align=center] at (7,1.5) {\lstinline{host.o}};

\draw[->] (5,-0.5) -- (6,-0.5);

\draw (6,-1) rectangle (8,0);
\node[align=center] at (7,-0.5) {\lstinline{device.img}};

\draw (3,-2.5) rectangle (5,-1.5);
\node[align=center] at (4,-2.0) {\lstinline{dev.rtl.bc}};

\draw[->] (4,-1.5) -- (4,-1);

\draw (9,0) rectangle (11,1);
\node[align=center] at (10,0.5) {Offload\\Bundler};

\draw[->] (8,1.5) -- (8.5,1.5) -- (8.5,0.5) -- (9,0.5);
\draw[->] (8,-0.5) -- (8.5,-0.5) -- (8.5,0.5) -- (9,0.5);

\draw[->] (10,0) -- (10,-1.5);

\draw (9,-2.5) rectangle (11,-1.5);
\node[align=center] at (10,-2) {\lstinline{foo.o}};

\end{tikzpicture}
\caption{Compilation flow of an OpenMP program with target offloading.}
\label{fig:omp-compilation}
\end{figure}

The compilation of an \OpenMP{} program with target offloading directives contains the following two passes (as shown in \autoref{fig:omp-compilation}):
\begin{description}
\item[Host Code Compilation.]
This pass includes the regular compilation of code for the host and \OpenMP{} offloading code recognition as preparation for the second pass.
Offloading regions are replaced by calls to the corresponding host runtime library functions (e.g. \lstinline{__tgt_target} for the directive \lstinline{target} in LLVM \OpenMP{}) with suitable arguments, such as the kernel function identifier, base pointers of each captured variables and the number of kernel function arguments.
In addition, a fallback host version of the kernel function will be emitted in case target offloading fails at runtime.
\item[Device Code Compilation.]
This pass utilizes the recognized \OpenMP{} target offload regions, as well as related functions and captured variables, and then emits target dependent device code.
This includes one entry kernel function per target region, global variables (potentially in different address spaces), and device functions, as well as some target dependent metadata.
As part of this compilation the \OpenMP{} device runtime library is linked into the user code as an LLVM bitcode library (\lstinline{dev.rtl.bc} in the \autoref{fig:omp-compilation}).
\end{description}

\noindent
In addition to the \lstinline{target} construct (as well as its combined variants), \OpenMP{} provides the \lstinline{declare target} directive which specifies that variables and functions are mapped onto a target device, and should hence be usable in device code.
The \lstinline{declare variant} directive can be used to specify a context, e.g., the compilation for a specific target, in which a specialized function variant should replace the base version.

\subsection{Motivation}
While the \OpenMP{} device runtime library can be implemented in any language it should be linked into the application in LLVM bitcode format for performance reasons.
This setup, shown in \autoref{fig:omp-compilation}, allows to optimize the runtime together with the application, effectively specializing a generic runtime as needed.

Given that the base language is irrelevant as long as we can compile to LLVM bitcode, \OpenMP{} comes to mind as a portable and performant way to write code for different accelerators.
As almost the entire device library can be interpreted as C++ code, rather than a CUDA or HIP code base, the compilation as \OpenMP{} is feasible,
in particular because LLVM/Clang is a working C++ and \OpenMP{} compiler already.

Since \OpenMP{} 5.1 all conceptually necessary building blocks are present in the language specification:

\begin{itemize}
\item The \lstinline{declare target} directive can be used to compile for a device, hence to generate LLVM bitcode that is targeting Nvidia's PTX or AMD's GCN.
As we do not need a host version at all, we can even use the LLVM/Clang flag \lstinline|-fopenmp-is-device| to invoke only the device compilation pass described in Section~\ref{sec:compilation-flow}.
\item The \lstinline{declare variant} directive can be used if a target requires a function implementation or global variable definition different from the default.
\item The \lstinline{allocate} directive provides access to the different kinds of memory on the GPU.
\end{itemize}

\noindent
For an additional target architecture, the work done in the compiler backend to emit code for that architecture will allow to retarget an \OpenMP{} implemented device runtime almost for free.
The incremental development cost is reduced from (re)implementing the device runtime in a language that can be compiled to the new architecture to
providing a few declare variant specialisations.

Finally, if the port uses compiler intrinsics instead of CUDA or HIP functions for the small target dependent part, it can be compiled without a vendor specific SDK present.
This unblocks shipping offloading as part of Linux distributions.

\section{Implementation}\label{sec:implementation}

In this section, we describe the new LLVM \OpenMP{} device runtime implemented with \OpenMP{} 5.1.
First, we talk about the common part, and then discuss how target dependent parts are implemented and why extensions were necessary.
Only AMD and Nvidia platforms are discussed as other GPU architectures cannot be targeted by the community LLVM version at this time.

\subsection{Common Part}
\paragraph{Device Code}~\\
Using the \lstinline{declare target} directive around all source files causes all functions and data to be emitted for the target device.
Macros to indicate that functions or globals are for the device, as shown in \autoref{lst:macros_in_devicertls}, are not needed.


\paragraph{Global Shared Variables}~\\
The implementation of the device runtime maps an \OpenMP{} team to a thread block\footnote{We are using CUDA terminology here. For AMD platforms it is \textit{wavefront}.} on the target device.
Therefore, a shared variable visible to all threads in the same thread block is equivalent to a variable that can be accessed within the same \OpenMP{} team.
The \lstinline{allocate} directive specifies how to allocate variables in different memory spaces.
Uses with an \lstinline{allocator(omp_cgroup_mem_alloc)}\footnote{The implementation currently uses \lstinline{allocator(omp_pteam_mem_alloc)} which is equivalent given the current mapping of parallelism.} we can place global variables in local shared memory, the equivalent of the CUDA \lstinline|__shared__| shown in \autoref{lst:macros_in_devicertls}.

In contrast to shared CUDA or HIP variables, C++ specifies that global variables are default initialized.
While we can technically do this for global shared variables defined with \OpenMP{}, it is not supported by LLVM/Clang at this time.
Furthermore, the performance is likely to suffer as the device runtime is designed to initialize these variables explicitly on demand.
To this end, we extended LLVM/Clang with a variable attribute for this work: \lstinline{loader_uninitialized} \cite{loader_uninitialized}.
The effect is that annotated variables will not have a default initialized value but instead be uninitialized like the CUDA or HIP shared variables are as well.

\autoref{lst:example_code} shows device code and global shared variable declaration as it is used in our \OpenMP{} device runtime.

\begin{lstlisting}[style=c_code_style, caption={An example of new device runtime code.}, label={lst:example_code}]
#pragma omp begin declare target

// Function declaration
extern __kmpc_impl_threadfence();
// Function definition
void __kmpc_flush(kmp_Ident *loc) {
  __kmpc_impl_threadfence();
}
// Global variable
int global_var;
// Shared variable
int shared_var;
#pragma omp allocate(shared_var).         \
            allocator(omp_pteam_mem_alloc)
// Shared variable declaration
extern int other_shared_var;
#pragma omp allocate(other_shared_var)    \
            allocator(omp_pteam_mem_alloc)
            
#pragma omp end declare target
\end{lstlisting}

\paragraph{Atomic Operations}~\\
The device runtime uses five atomic operations, \lstinline{add}, \lstinline{inc}, \lstinline{max}, \lstinline{exchange}, and \lstinline{cas}, implemented in target dependent parts with LLVM/Clang builtin functions.

\OpenMP{} 5.1 \cite{openmp-5.1-spec} introduces the \lstinline{compare} clause, which supports conditional update statements.
When combined with the \lstinline{capture} clause, all of these atomic operations except \lstinline{inc} can be implemented via \OpenMP{}, as shown in \autoref{lst:atomic_operations}.
We implemented the support of the \lstinline{compare} clause and its combination with the \lstinline{capture} clause for LLVM/Clang but the it has not been merged into the community version yet.
With the updated requirements for flush\footnote{\OpenMP{} 5.1 removes the requirement for a flush operation at the entry and exit of an atomic operation if \lstinline{write}, \lstinline{update}, or \lstinline{capture} is specified and the memory ordering is \lstinline{seq_cst}},
which we also implemented for this work,
our \OpenMP{} versions of atomic operations can generate LLVM-IR that is identical to the original target dependent implementation via compiler intrinsics.

\begin{lstlisting}[style=c_code_style, caption={Atomic operations implemented in \OpenMP{} 5.1.}, label={lst:atomic_operations}]
uint32_t atomic_add(uint32_t *X, uint32_t E) {
  uint32_t V;
#pragma omp atomic capture seq_cst
  { V = *X; *X += E; }
  return V;
}
uint32_t atomic_max(uint32_t *X, uint32_t E) {
  uint32_t V;
#pragma omp atomic compare capture seq_cst
  { V = *X; if (*X < E) { *X = E; } }
  return V;
}
uint32_t atomic_exchange(uint32_t *X, uint32_t E) {
  uint32_t V;
#pragma omp atomic capture seq_cst
  { V = *X; *X = E; }
  return V;
}
uint32_t atomic_cas(uint32_t *X, uint32_t E, uint32_t D) {
  uint32_t V;
#pragma omp atomic compare capture seq_cst
  { V = *X; if (*X == E) { *X = D; } }
  return V;
}
\end{lstlisting}

\noindent
The missing atomic operation is \lstinline|inc|.
According to the CUDA specification~\cite{cuda-doc}, \lstinline{inc} implements:
\begin{lstlisting}[style=c_code_style]
{ v = x; x = x >= e ? 0 : x + 1; }
\end{lstlisting}
and returns \lstinline{v}.
This atomic operation can not be represented in a form that \OpenMP{} 5.1 supports because \OpenMP{} 5.1 requires that the order operation be either \lstinline{<} or \lstinline{>}, and the alternative statement of the conditional expression statement must be \lstinline{x} itself.
Therefore, we still keep it in the target dependent part implemented with LLVM intrinsics as shown in \autoref{lst:atomic_inc}.

\subsection{Target Specific Part}
Target dependent global functions and variables are currently declared in a header and implemented in target dependent source files which are only compiled for the specific target, either as CUDA or HIP.
A drawback of this method is that the creation of a device runtime for a new target might require us to remove a function from the common part and insert it into the target specific part if the existing (common) implementation is not suited for the new device.

Since \OpenMP{} 5.0, the \lstinline{declare variant} directive declares a specialized variant of a base function and specifies the context in which that specialized variant is used.
It supports various context selector with the \lstinline{match} clause, one of which is \lstinline{device} selector.
For example, with \lstinline|match(device={arch(arch_name)})|, the code wrapped in a \lstinline{begin/end declare variant} region will be only generated if the target architecture \textit{matches} the \lstinline{arch_name}.

\autoref{lst:atomic_inc} shows how the atomic \lstinline{inc} function is implemented with target dependent compiler intrinsics selected via the \lstinline|begin/end declare variant| directive for both Nvidia and AMD GPU targets.

Note that we use the \lstinline{match_any} extension for Nvidia platforms as we support two distinct architectures, \lstinline{nvptx} and \lstinline{nvptx64}, but we do not want to distinguish between them in the device runtime.
While this can be handled by duplicating the code, our new context selector
changes the semantic of the matching to produce a match if \textit{any} architecture in \lstinline{arch(nvptx, nvptx64)} is targeted.
By default a match would require all architectures to be targeted.
In addition to \lstinline|match_any| we extended LLVM/Clang with other useful context selectors, e.g, \lstinline|match_none| and \lstinline|allow_templates|\footnote{See: \url{https://clang.llvm.org/docs/AttributeReference.html\#pragma-omp-declare-variant}}.

\begin{lstlisting}[style=c_code_style, caption={Atomic \lstinline{inc} implementation with the \lstinline{match_any} clause.}, label={lst:atomic_inc}]
#pragma omp declare target
// Fallback version, which raises a compilation error
uint32_t atomic_inc(uint32_t *X, uint32_t E) {
  error("target dependent implementation missing");
}
// AMDGCN implementation
#pragma omp begin declare variant                       \
            match(device={arch(amdgcn)})
uint32_t atomic_inc(uint32_t *X, uint32_t E) {
  return __builtin_amdgcn_atomic_inc32(X, E,
                                       __ATOMIC_SEQ_CST, "");
}
#pragma omp end declare variant
// NVPTX implementation
#pragma omp begin declare variant                       \
            match(device={arch(nvptx,nvptx64)},         \
                  implementation={extension(match_any)})
uint32_t atomic_inc(uint32_t *X, uint32_t E) {
  return __nvvm_atom_inc_gen_ui(X, E);
}
#pragma omp end declare variant
#pragma omp end declare target
\end{lstlisting}

\noindent
Other target dependent functions are required to handle synchronization, thread hierarchy, etc.
These are implemented via compiler intrinsics, function calls to the corresponding native runtime library, or inline assembly.

%
%
%

\section{Evaluation}\label{sec:evaluation}
In this section, we evaluated our proposed method in three ways: code comparison, functional testing, and performance evaluation.

\subsection{Code Comparison}
The previous implementation compiled CUDA to LLVM-IR, and HIP to LLVM-IR, while our proposed method compiles \OpenMP{} to LLVM-IR for both platforms.
The accuracy of the port to \OpenMP{} was assessed by comparing the text form of the library before and after changing over to \OpenMP{}.
If the text forms were identical, we would be certain the language change made no difference.
This was not quite the case.
The differences were in semantically unimportant metadata, symbol name mangling for variant functions, and the order of inlining as preferred by the language front end which had minor reordering effects on PTX and GCN generation.

\subsection{Functional Testing}
There are a number of \OpenMP{} test suites and applications in use for checking the behaviour of the compiler, including SOLLVE V\&V \cite{sollvevv}, and Ovo \cite{ovo}.
All ran identically with the new \OpenMP{} runtime as they had using the previous device runtime.

\subsection{Performance Evaluation}
\subsubsection{Systems Configuration}
We evaluate the performance of our method experimentally on the \Summit{} supercomputer.
Each \Summit{} node contains two IBM POWER9 processors and six Nvidia Volta V100 GPUs.
CUDA 10.1.243 was used, which is the version loaded by default.

\subsubsection{Benchmarks}
The SPEC ACCEL benchmark suite V1.3 was used to evaluate the new device runtime.
Because support for Fortran is still in progress, we chose those benchmarks written in C.
There are 15 \OpenMP{} enabled benchmarks in SPEC ACCEL. Seven of them are in C, namely  \lstinline{503.postencil}, \lstinline{504.polbm}, \lstinline{514.pomriq}, \lstinline{552.pep}, \lstinline{554.pcg}, \lstinline{557.pcsp}, and \lstinline{570.pbt}.
\lstinline{557.pcsp} can not be compiled, therefore we only ran the other six benchmarks.
We also chose a  C++ proxy application, miniQMC \cite{qmcpack}.

\lstinline{O2} was used when compiling the benchmarks and application.
Each test case was executed five times, and the execution time was averaged.
miniQMC was measured through the \lstinline|miniqmc_sync_move| benchmark executed
as follows: \lstinline{miniqmc_sync_move -g "2 2 1"}.

\subsubsection{Results}
\begin{figure}[hbt!]
\centering
\vspace*{-3mm}
\begin{tikzpicture}

\begin{axis}  
[
width=0.8\textwidth,
height=0.4\textwidth,
ybar,
ylabel={Execution Time (s)},
ymin=0, ymax=80,
symbolic x coords={503.postencil, 504.polbm, 514.pomriq, 552.pep, 554.pcg, 570.pbt},  
xtick=data,
ytick={10, 20, 30, 40, 50, 60, 70},
x tick label style={font=\small, rotate=30},
nodes near coords,
visualization depends on={rawy \as \rawy},
nodes near coords={\pgfmathprintnumber\rawy},
restrict y to domain*={
\pgfkeysvalueof{/pgfplots/ymin}:\pgfkeysvalueof{/pgfplots/ymax}
},
every node near coord/.append style={font=\tiny},
nodes near coords align={vertical},
]

\addplot[fill=none] coordinates {(503.postencil, 13.8) (504.polbm, 32.2) (514.pomriq, 28.2) (552.pep, 68.6) (554.pcg, 98.1) (570.pbt, 513)};

\addplot[pattern=north east lines] coordinates {(503.postencil, 13.6) (504.polbm, 32.2) (514.pomriq, 28.2) (552.pep, 68.6) (554.pcg, 98.1) (570.pbt, 514)};

\end{axis}

\end{tikzpicture}
\vspace*{-5mm}
\caption{Comparison between execution time of original device runtime ({\protect\tikz\protect\draw(0,0) rectangle (0.3,0.2);}) and that of our proposed new device runtime ({\protect\tikz\protect\draw[pattern=north east lines] (0,0) rectangle (0.3,0.2);}) on Nvidia platform.}
\label{fig:spec_nvidia}
\vspace*{-6mm}
\end{figure}
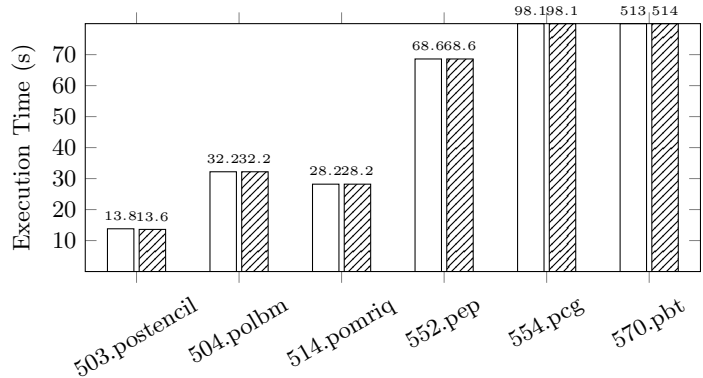

\autoref{fig:spec_nvidia} 
compares the execution time when the original device runtime is used with the execution time obtained using 
our proposed new device runtime. 
We can see that the execution times are almost identical, and for those cases where they are not same, the variance is less than $1\%$ and assumed to be noise.

The proxy application benchmark \lstinline{miniqmc_sync_move} contains two target regions, \lstinline{evaluate_vgh} and \lstinline{evaluateDetRatios}.
They are executed  multiple times.
\autoref{tbl:miniqmc_nvidia} shows the profiling results (execution time) of each target region from Nvidia's profiler \lstinline{nvprof}.
There is no performance difference between the two versions.

\begin{table}[hbt!]
\vspace*{-2mm}
\centering
\begin{tabular}{|c|c|r|c|r|r|r|}
\hline
Target Region &
  Version &
  \multicolumn{1}{c|}{Time (ms)} &
  \multicolumn{1}{c|}{\# Calls} &
  \multicolumn{1}{c|}{Avg ($\mu$s)} &
  \multicolumn{1}{c|}{Min ($\mu$s)} &
  \multicolumn{1}{c|}{Max ($\mu$s)} \\ \hline
\multirow{2}{*}{\lstinline{evaluate_vgh}}      & Original & 1374.72 & \multirow{2}{*}{64512} & 21.309 & 19.744 & 32.384 \\ \cline{2-3} \cline{5-7} 
                                               & New      & 1376.59 &                        & 21.338 & 19.776 & 33.760 \\ \hline
\multirow{2}{*}{\lstinline{evaluateDetRatios}} & Orignal  & 573.46  & \multirow{2}{*}{18202} & 31.505 & 25.247 & 44.480 \\ \cline{2-3} \cline{5-7} 
                                               & New      & 573.93  &                        & 31.531 & 24.544 & 47.103 \\ \hline
\end{tabular}
\vspace*{2mm}
\caption{Comparison of execution time of the two target regions in \lstinline{miniqmc_sync_move} on Nvidia platform.}
\label{tbl:miniqmc_nvidia}
\vspace*{-6mm}
\end{table}

All the results above demonstrate that our proposed portable \OpenMP{} device runtime can provide the same performance as the current CUDA-like version on the Nvidia platform. Based on the code comparison, functional testing and some AMD internal performance testing results, the portable runtime is believed to show no performance change from its HIP predecessor either.


\section{Conclusions and Future Work}\label{sec:conclusion}
\OpenMP{} works well as a language to implementing GPU-only code libraries.
The direct support for memory allocators and the precise dispatch through \lstinline{declare variant} are clear advantages over C++.
While minimal compiler modifications were required to match the CUDA and HIP semantics to the fullest, we expect those to be incorporated into the \OpenMP{} standard over time.

Using \OpenMP{} is especially suitable as the vehicle for implementing an \OpenMP{} runtime library since the main prerequisite is an \OpenMP{} compiler which needs to be implemented all targets in any case.
Since the library ships with the LLVM repository, it can be built by any distribution which has built Clang.
Vendor SDKs or compilers are no longer required. 

Since the host and device runtime libraries can build as part of LLVM, we will coordinate with Linux distribution developers to ensure that people who install the distribution LLVM package onto a system that has a target device and driver available will be able to get this  working ``out of the box''.

\section*{Acknowledgement}
This research was supported by the Exascale Computing Project (17-SC-20-SC), a collaborative effort of two U.S. Department of Energy organizations (Office of Science and the National Nuclear Security Administration) responsible for the planning and preparation of a capable exascale ecosystem, including software, applications, hardware, advanced system engineering, and early testbed platforms, in support of the nation's exascale computing imperative.

%
%
\bibliographystyle{splncs04}
\bibliography{reference.bib}

\end{document}